\documentstyle[prb,multicol,amsmath,latexsym,float,epsfig,aps]{revtex}

\def\gee	{\epsilon}

\def\go		{\omega}

\def\capo	{\right.\\ \left.}

\def\la		{\langle}
\def\ra		{\rangle}

\renewcommand{\[}{\left[}
\renewcommand{\]}{\right]}
\renewcommand{\(}{\left(}
\renewcommand{\)}{\right)}

\restylefloat{figure}

\begin{document}
\draft
\title{Plane-waves DFT-LDA calculation of the
 electronic structure and absorption spectrum of Copper}
\author{
Andrea Marini, Giovanni Onida and Rodolfo Del Sole}
\address{
Istituto Nazionale per la Fisica della Materia, Dipartimento di Fisica
   dell'Universit\`a \\ di Roma ``Tor Vergata'',
   Via della Ricerca Scientifica, I--00133 Roma,
   Italy }
\date{\today}
\maketitle

\begin{abstract}
We present an accurate, first-principles study of the electronic structure and 
absorption spectrum of bulk copper within Density Functional Theory in the 
Local Density Approximation (DFT-LDA), including the study  of intraband transitions.
We construct norm-conserving pseudopotentials (PPs) including the 3d shell 
(and optionally the underlying 3s and 3p shells) in the valence, 
and requiring a relatively small plane-waves basis (60 and 140 Rydbergs cutoff, respectively). 
As a consequence, these PPs are strongly non-local, yielding to
macroscopically wrong results in the absorption spectrum when momentum matrix elements 
are computed naively. 
Our results are compared with experimental photoemission, 
absorption and electron energy loss data, and suggest non trivial self-energy effects in 
the quasiparticle spectrum of Cu.
\end{abstract}


\begin{multicols}{2}
\narrowtext

%%%%%%%%%%%%%%%%%%%%%%%%
\section{Introduction}
%%%%%%%%%%%%%%%%%%%%%%%%
Copper has played for a long time a central role in the elucidation of the electronic
structure of solids. It is a relatively inert material, very easy to handle
experimentally; its
energy dispersion relations (including spin-orbit interaction for some bands) have been
measured with considerable precision; lifetimes of the band states as a function of the 
distance from Fermi level have been determined; surface states have been analyzed in
various parts of Surface Brillouin zone (for a review see\cite{courths}).
From the theoretical side, the
study of noble metals like Copper using first-principles
methods based on plane-waves and ab-initio pseudopotentials (PPs) presents some peculiar
complication with respect to the case of simple metals or semiconductors. 
In fact, in addition to metalicity, which implies the use of an accurate sampling of 
the Brillouin zone in order to describe properly the Fermi surface, one must also keep 
into account the contribution of d-electrons to the bonding and to 
the valence bandstructure. 
This means that, within the PP scheme, d states cannot be 
frozen into the core part, but must be explicitly included into the valence, 
yielding a large total number of valence electrons (11 for bulk 
copper).
Unfortunately, a Cu pseudopotential including 3$^{rd}$ shell states
into the valence part is very steep. Hence, when working with a
plane--wave basis,
the usage of a PP of this kind may be computationally prohibitive.

On the other hand, the use of a pseudopotential without explicit treatment of
3d-electrons has been shown to be unreliable \cite{campillo}. 
Hence, first-principles methods based on plane-waves have been  
used only seldom to treat Cu \cite{campillo}.
However, methods have been devised 
for the construction of softer pseudopotentials 
\cite{vdb,rappe,shir,TM}, which make the inclusion of 
the  3$^{rd}$ shell in the valence more affordable. 
The price to be paid is that the construction and use of such 
pseudopotentials, which often
display a very strong l-nonlocality, 
is a quite delicate matter. In particular, 
the choice of  a reference 
component \cite{BHS} and the transferability checks must be done with care. 
These difficulties are more than compensated by the simplicity 
and elegance of the 
plane-waves formalism in the subsequent calculations. 
In the present work, two possible choices for the Cu pseudopotential 
are explored, i.e: a) including 3s and 3p electrons into the 
frozen core (a quite standard choice), and b) including the full
3$^{rd}$ shell in the valence. Fully converged 
Density Functional Theory -- Local Density Approximation (DFT-LDA)
calculations are performed at 60 and 140 Ry cutoff, respectively 
for case a) and b). 

The paper is organized as follows: in Section \ref{sec:PP} we give the details of the 
construction of the pseudopotentials used 
throughout this work ; in Sections \ref{sec:GS} and
\ref{sec:lda_bands} we present the ground state properties and bandstructure, 
respectively, obtained with the different pseudopotentials;
finally, in Section \ref{sec:e2}, we compare the theoretical absorption and electron 
energy loss spectra
with the experimental data, including the 
effects of Local Fields  and intraband transitions.

%%%%%%%%%%%%%%%%%%%%%%%%%%%%%%%%%%%%%%
\section{Pseudopotential generation}
\label{sec:PP}
%%%%%%%%%%%%%%%%%%%%%%%%%%%%%%%%%%%%%%

The Cu atom has the ground--state electronic configuration [Ar]$3d^{10}4s^1$
(or [Ne]$3s^23p^63d^{10}4s^1$). 4s and 3d eigenvalues 
are separated, in DFT--LDA,
by less than 0.5 eV. It is then quite obvious that freezing all states but the 
4s one into the atomic core (i.e., to neglect the polarization of the 3d 
electrons) cannot yield a good, transferable pseudopotential.
On the other hand, inclusion of the 3d electrons into the valence 
(we call this a ``3d'' pseudopotential) 
yields a much slower convergence of plane--wave expansions, due to 
the steepness of the d-component of the PP. 
Unfortunately, the 
spatial superposition between the 3d and 3s or 3p states is quite large,
despite the large ($\simeq 70$ eV) energy separation.
Hence, an even more conservative and secure choice for the PP 
is to include 
all 3s, 3p and 3d electrons into the valence: 
in fact, 2p and 3s states  
are well separated, both spatially and energetically ($\simeq$ 800 eV).
This latter choice gives rise to a PP which is even harder than the ``3d''
one, and which will be referred to as a ``3s'' pseudopotential, yielding
19 valence electrons per atom.

Using the traditional  pseudopotential generation scheme proposed by 
Bachelet, Hamann and Schl\"{u}ter
(BHS) \cite{BHS} yields PP whose d component converges very
slowly in Fourier space, requiring to work at an energy 
cutoff of 200 rydbergs or more.
However, PPs which converge at less than 100 Rydbergs
can be constructed by using  specially devised 
schemes as those of refs. \cite{vdb,rappe,shir,TM}.
We choose to restrict to the class of {\it norm conserving}
pseudopotentials, in order to avoid the additional numerical 
complications
arising from the charge-state dependence of the PP of
ref. \cite{vdb}. 
Moreover, our choices in generating the PP are dictated not only
by the need of a fast convergence of the PP in Fourier space,
but also by that of optimizing the PP accuracy and transferability,
a non--trivial task when also 3s and 3p states are included into
the valence.
In particular, the Hamann scheme \cite{hamann} (which
does not try to optimize the Fourier space convergence at all)
turned out to yield much more accurate and transferrable 
norm--conserving PPs,
particularly when the whole 3$^{rd}$ shell is included into the 
valence.  

Hence, we strictly follow the Hamann procedure (described in the 
appendix of ref. \cite{hamann}) whenever it is possible,
i.e. in all cases except for the d components. For the latter,
we follow the
Troullier--Martins \cite {TM} scheme, which  allows us to 
reduce significantly
the number of plane--waves requested for convergence without loosing
too much in transferability.
In all other cases (i.e. for the s and p
components of both ``3s'' and ``3d'' pseudopotentials) we found the
Hamann procedure to be more convenient, even at the cost of a slower
Fourier space convergence, since the TM one yielded 
significantly worse results, and/or ghost states \cite{gonze_ghost}
when the PP were used in the Kleinman--Bylander form \cite{KB}.

Another delicate point is the choice of a reference component, i.e.
of a PP angular momentum  component  which is taken to be valid 
for every l$\geq 3$. Often, the l=2 component is chosen as a reference,
simply because this makes calculations easier. This choice is 
sometimes lacking a physical justification, and can be dangerous,
as it has been shown by some of us in the case of Sn \cite{sno}. 
In the present case the l=2 reference had to be avoided anyway,
since it yielded a much worse transferability than the l=0 or l=1
choices, and sometimes gave rise to ghost states in the KB form.

Several trials and tests with different cutoff radii have been done, in 
order to optimize the PP transferability without increasing too much
the number of plane waves requested for convergence. Transferability tests
included both the plot of logarithmic derivatives, and the explicit 
calculation of pseudoatom eigenvalues in some excited electronic 
configuration (both neutral and positively charged).

Finally, since
3s and 3p states in the solid preserve their atomic configuration 
better than 3d ones, 
their explicit inclusion into the valence can sometimes be avoided, 
by considering only 
the effects of the non-linearity of the
exchange-correlation potential \cite{NLCC}.
Hence, a  third pseudopotential, with frozen 3s and 3p electrons 
but including  non-linear 
core-corrections has also been considered, and will be referred to as
the ``3d+NLCC'' PP. In the latter, the core charge is represented by the
true one for r$\geq 0.5$ Bohrs, and by a Gaussian model charge for 
r$\leq 0.5$ Bohrs.

The resulting optimal cutoff radii and reference l--components,
chosen for our ``3d'', ``3s'', and ``3d+NLCC'' pseudopotentials, are given in
Table \ref{table_rc_pseu}. 

\begin{figure}[H]
\begin{center}
\epsfig{figure=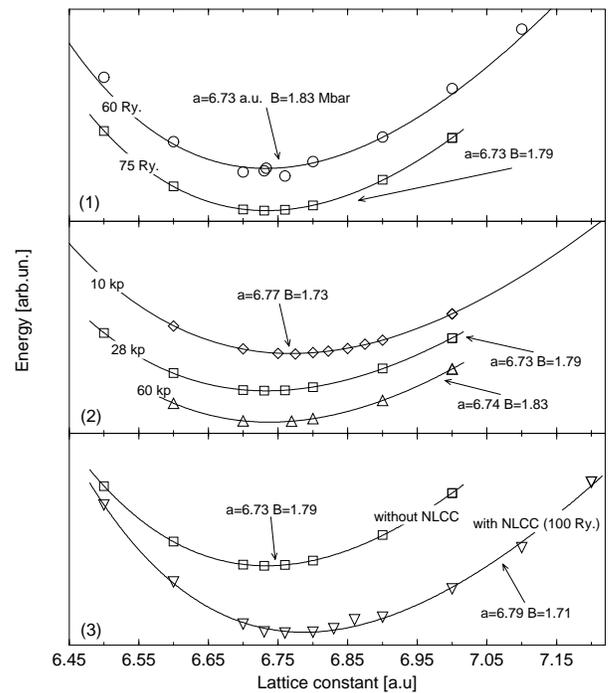,width=8cm}
\end{center}
\caption {\footnotesize{
Calculated total energy vs. lattice constant for bulk Cu.
Panel (1): effect of the kinetic energy cutoff at fixed number
of k--points (28).
Panel (2): effect of the IBZ sampling at fixed $E_{cut}$
(75 Ry.);
Panel (3): effect of nonlinear core--corrections  (see text).
The reported values for the equilibrium lattice constant
and bulk modulus have been obtained from Murnaghan fits (continuous curves).
The experimental values are a=6.822 Bohrs  and B=1.827 MBar  {\protect  \cite{expt}}. }}
\label{fig:ground1}
\end{figure}

%%%%%%%%%%%%%%%%%%%%%%%%%%%%%%%%%
\section{Ground state properties}
\label{sec:GS}
%%%%%%%%%%%%%%%%%%%%%%%%%%%%%%%%%

Our first step is a self--consistent ground--state calculation, performed by
minimizing the DFT--LDA energy functional with  a Car--Parrinello
method\cite{CP}, in a standard plane--wave basis.
The Ceperley--Alder\cite{cepal} exchange--correlation energy and potential,
as parametrized by Perdew and Zunger\cite{perzun}, have been used (test 
calculations with the Hedin--Lundqvist form \cite{hedin_lund} 
have also been performed: see below).
All our pseudopotentials are used within the fully--separable Kleinman--Bylander scheme
\cite{KB}, after checking that no ghost--states was present \cite{gonze_ghost}.
The irreducible wedge of the Brillouin zone (IBZ) was sampled with the use
of Monkhorst and Pack (MP) sets\cite{MP} of $N_k$ k--points.
A very small fictitious electronic
temperature (equal to $\approx$ 10 Kelvin) was used in order to accelerate the 
convergence of the calculated Fermi surface.
Convergence with respect to both the k--points sampling and the kinetic energy cutoff
has been checked extensively: Fig. \ref{fig:ground1} and Table \ref{table_conv} show 
the results obtained for the ``3d'' pseudopotential.
E$_{cut}$ = 60 Ry. and N$_k$ = 28 appear to give well--converged, satisfactory results,
with lattice constant a$_0$ and Bulk modulus B$_0$ within 1.4$\%$ of the experimental values.
In the calculation with NLCC, the energy cutoff had to be increased to 100 Ry,
in order to describe properly the core charge.  NLCC  reduce 
the underestimation of the experimental lattice constant to 0.5 $\%$,
but they are found to be almost ininfluent on the bandstructure, as well as on the 
resulting spectra (see below). 
Finally, calculations with the ``3s'' pseudopotential, much deeper than the ``3d''
one,  required an energy cutoff of 140 Ry. Also in this case,  the effect on the
LDA bandstructure and spectra in the energetic  region of interest 
are found to be very small (see below). 

In summary, as long as only the LDA 4s and 3d bandstructure is concerned,
the effect of the inclusion of 3s and 3p in the valence, (as well as the use of 
NLCC), are mainly confined to a change of the equilibrium lattice constant, 
which induces an indirect effect on the bandstructure energies \cite{nota1}. 

\begin{figure}[H]
\begin{center}
\epsfig{figure=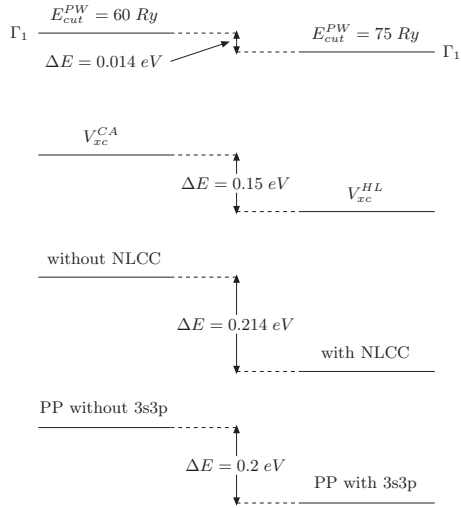,width=6cm}
\end{center}
\caption {\footnotesize{
Summary of the effects of different computational details on the
LDA bandstructure. The energy differences reported are the maximum ones,
and correspond to the bottom valence band at the
$\Gamma$ point ($\Gamma_1$).
They decrease gradually to zero at the  Fermi level.}}
\label{fig:lda_bands.2}
\end{figure}

%%%%%%%%%%%%%%%%%%%%%%%%%%%%
\section{LDA band structure}
\label{sec:lda_bands}
%%%%%%%%%%%%%%%%%%%%%%%%%%%%

Fig. \ref{fig:lda_bands.2} summarizes the maximum relative energy differences 
induced by different computational details on the LDA bandstructure;
the value is taken at the 
bottom valence in the $\Gamma$ point ($\Gamma_1$) and decreases gradually 
to zero at the Fermi level.

By changing $E_{cut}$ from 60 Ry to 75 Ry the band-structure 
remains practically identical (changes are less then 0.02 meV).
At the  60 Ry cutoff, we compare the band structure calculated with the
Ceperley-Alder parametrization of the exchange-correlation potential \cite{cepal,perzun} with that
obtained using the Hedin-Lunqvist parametrization
\cite{hedin_lund}; in the latter case we find a maximum energy shift of about 0.15 eV.

\begin{figure}[H]
\begin{center}
\epsfig{figure=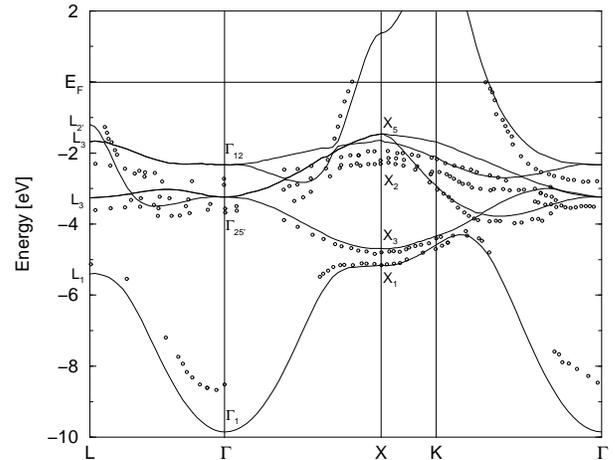,width=8cm}
\end{center}
\caption{\footnotesize{Bulk copper DFT-LDA bandstructure (full--line),
compared with photoemission
Data (points) from Ref. {\protect\cite{courths}}.}}
\label{fig:lda_bands.1}
\end{figure}

The inclusion of NLCC yields, instead, a maximum upward
shift of about 0.21 eV with respect to the case without NLCC, mainly due to the change in the
equilibrium lattice constant.
To study more deeply the effects of core polarization, we also performed a band-structure calculation
with the ``3s'' pseudopotential (see Section \ref{sec:PP}), where the
3s and 3p core level relaxation is fully included in the selfconsistent run.
Also in this case the maximum bandshift with respect to the ``3d'' pseudopotential is limited to
about 0.2 eV.
Hence, our results obtained with the ``3d'' pseudopotential at  $E_{cut}$=60Ry. and
using 28 MP k--points in the determination of the self--consistent charge density
can be considered to represent the converged LDA bandstructure of bulk copper.
Theoretical results are compared
with the experimental photoemission data in Fig. \ref{fig:lda_bands.1}.
At difference with the case of semiconductors, the disagreement between
theory and experiment is far from being limited to a rigid shift of the 
Kohn-Sham occupied eigenvalues with respect to the empty ones.
In fact, as also summarized in Table \ref{tab:lda_bands.1}, the widths of the 
d bands are systematically overestimated, a well-known failure of LDA when applied to
transition and noble metals\cite{gunn_rev}.

%%%%%%%%%%%%%%%%%%%%%%%%%%%%%
\section{Absorption spectrum}
\label{sec:e2}
%%%%%%%%%%%%%%%%%%%%%%%%%%%%%

The absorption spectrum is given by the imaginary part of the macroscopic
dielectric function:
\begin{align}
\gee_M\(\go\)=\frac{1}{\gee^{-1}_{{\bf 0\;0}}\(\go\)},\label{eq:e2.1}
\end{align}
where
\begin{align}
 \gee^{-1}_{{\bf 0\;0}}\(\go\)=1+\lim_{{\bf q}\rightarrow {\bf 0}}
 \frac{4 \pi}{|{\bf q}|^2}\chi_{{\bf G}={\bf 0}\;{\bf G}'={\bf 0}}\({\bf q},\go\),\label{eq:e2.2}
\end{align}
$\chi_{{\bf G\;G}'}\({\bf q},\go\)$ is the reducible polarization, solution of the equation
\begin{multline}
 \chi_{{\bf G\;G'}}\({\bf q},\go\)=\chi^0_{{\bf G\;G'}}\({\bf q},\go\)+\\
 \sum_{{\bf  G''}}\chi^0_{{\bf G\;G''}}\({\bf q},\go\)
 \frac{4 \pi}{|{\bf q+G''}|^2}\chi_{{\bf G''\;G'}}\({\bf q},\go\).\label{eq:e2.3}
\end{multline}
Neglecting (for the moment) intraband transitions, $\chi^0_{{\bf G\;G''}}\({\bf q},\go\)$
is given by:
\begin{multline}
 \chi^0_{{\bf G\;G'}}\({\bf q},\go\)=\frac{1}{2} 
 \int_{BZ} \frac{d^3 {\bf k}}{\(2 \pi\)^3}\\
 \sum_{n\neq n'}
 \la n'{\bf k-q}|e^{-i\({\bf q+G}\)\cdot{\bf r}}|n {\bf k}\ra
 \la n{\bf k}|e^{i\({\bf q+G'}\)\cdot{\bf r}'}|n' {\bf k-q}\ra\\
 {\cal G}^0_{he}\(n,n',{\bf k},{\bf q},\go\),\label{eq:e2.4}
\end{multline}
with
\begin{multline}
{\cal G}^0_{he}\(n,n',{\bf k},{\bf q},\go\)=f_{n'}\({\bf k-q}\)\(2-f_n\({\bf k}\)\) \\
 \[\frac{1}{\go+\gee_{n'}\({\bf k-q}\)-\gee_n\({\bf k}\)+i\eta}-\capo
 \frac{1}{\go+\gee_n\({\bf k}\)-\gee_{n'}\({\bf k-q}\)-i\eta}\].
 \label{eq:e2.5}
\end{multline}
with $0 \leq f_n\({\bf k}\)\leq 2$ representing the occupation number summed over 
spin components.
The sums over $k$ are transformed to integrals over the BZ, and the latter are evaluated by 
summing over large sets of random points contained in the whole BZ. 
Fully converged calculations with a small broadening require a very large number of 
k-points \cite{REPLY_ALBRECHT}; in the present work, the broadening used (and the
corresponding number of k--points which was found to be sufficient 
to ensure convergence) is specified explicitly for each one of the reported spectra.

The simplest approach to the calculation of  the absorption spectrum neglects the full
inversion of Eq. (\ref{eq:e2.3}) (i.e., neglects Local Field Effects), and assumes
\begin{align}
\gee_M\(\go\)\approx 1-\lim_{{\bf q}\rightarrow {\bf 0}}
 \frac{4 \pi}{|{\bf q}|^2}\chi_{{\bf G}={\bf 0}\;{\bf G}'={\bf 0}}\({\bf q},\go\).\label{eq:e2.6}
\end{align}
The ${\bf q}\rightarrow {\bf 0}$ limit for the oscillator strengths appearing in
Eq. (\ref{eq:e2.4}) is calculated 
in the transversal gauge \cite{rodolfo1}, within first order perturbation theory \cite{hybertsen1}
\begin{multline}
 \lim_{{\bf q}\rightarrow {\bf 0}}
 \la n'{\bf k-q}|e^{-i{\bf q}\cdot{\bf r}}|n {\bf k}\ra=\\
 -i{\bf q}\cdot\frac{\la \phi_{n'{\bf k-q}}|\[{\bf r},H\]|\phi_{n {\bf k}}\ra}
 {\gee_{n'}\({\bf k}\)-\gee_n\({\bf k}\)}+O\(q^2\),
 \label{eq:e2.7}
\end{multline}
where $\phi_{n{\bf k}}\({\bf r}\)$ are the Bloch functions.
Due to the non local character of the norm--conserving pseudopotentials,
the well-known relation between $\[{\bf r},H\]$ and the momentum operator
\begin{align}
 \[{\bf r},H\]={\bf p}.
 \label{eq:e2.9a}
\end{align}
must be substituted by:
\begin{align}
 \[{\bf r},H\]={\bf p}+\[{\bf r},V_{NL}\].
 \label{eq:e2.8}
\end{align}
The second term of the rhs of Eq. (\ref{eq:e2.8}), which in simple metals and in many
semiconductors is small (and often neglected in practical calculations), becomes extremely
important in the case of copper due to the large nonlocality of the PP.
\begin{figure}[H]
\begin{center}
\epsfig{figure=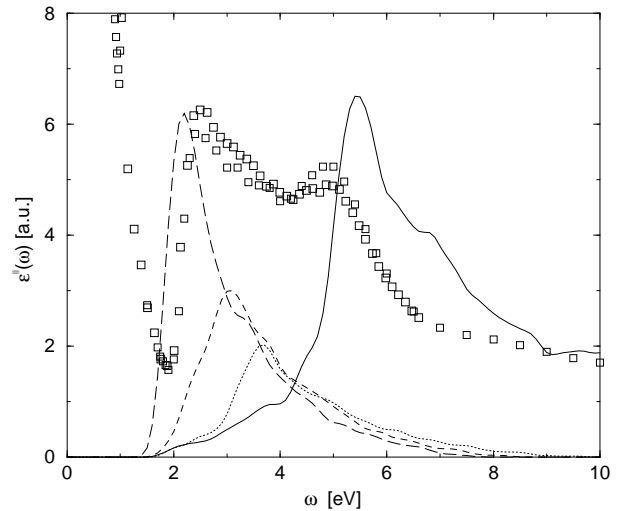,width=8cm}
\end{center}
\caption {\footnotesize{Full line: imaginary part of the
macroscopic dielectric function of Cu without Local-Field effects and without including the
non-local pseudopotential commutator [Eq. ({\protect \ref{eq:e2.8}})], compared with
experimental data (squares) from{\protect\cite{palik}}.
Dot, dashed and long--dashed lines correspond to the functions defined in Eq.
({\protect \ref{eq:e2.10}})
with $n'=3$, $n'=4$ and $n'=5$ respectively. All theoretical curves are computed with
N$_k=15,386$, and a Gaussian broadening of 0.15 eV (see text).}}
\label{fig:e2.1}
\end{figure}
This is demonstrated in Fig. \ref{fig:e2.1} where the imaginary part of
$\varepsilon_M$ is calculated assuming the validity of Eq. (\ref{eq:e2.9a})
The experimental absorption spectrum is severely underestimated 
between the offset of interband transitions ($\approx\;1.74$ eV) and $5$ eV.
A better analysis of this behavior can be performed by  plotting 
the quantity
\begin{align}
 \Im m\[\frac{\sum_{{\bf k}}{\cal G}^0_{he}\(n,n',{\bf k},{\bf 0},\go\)}{\go^2}\].
\label{eq:e2.10}
\end{align}
with $n=6$ and $n'=3,4,5$ (see Fig. \ref{fig:e2.1}).
This quantity is the joint density of states (JDOS) (divided by $\go^2$)  for transitions between
the sixth band (d--like) and the third to fifth bands (sp--like), and 
corresponds to assuming an oscillator strength equal to one in Eq. (\ref{eq:e2.4}).
\begin{figure}[H]
\begin{center}
\epsfig{figure=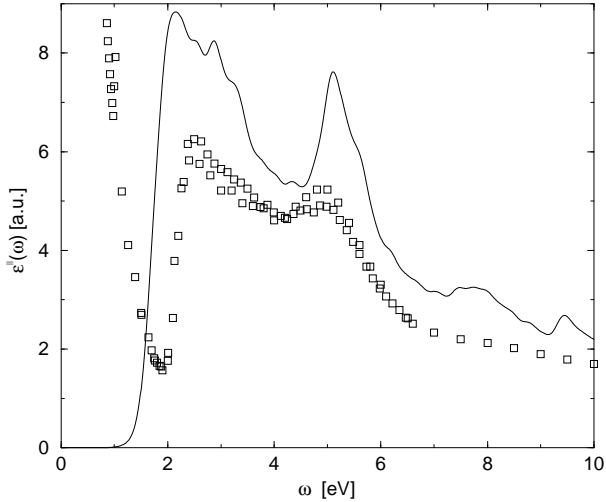,width=8cm}
\end{center}
\caption {\footnotesize{Full line: imaginary part of the
macroscopic dielectric function of Cu without Local-Field effects and
{\it including} the
non-local pseudopotential commutator [Eq. (\ref{eq:e2.8})], compared with
experimental data (squares){\protect\cite{palik}}.N$_k$ and  broadening
as in Fig. {\protect\ref{fig:e2.1}}.}}
\label{fig:e2.2}
\end{figure}

We see that transitions in the energy range of interest (1.8 to 5.0 eV) exist, but they are
strongly suppressed due to the small values of the corresponding matrix elements of {\bf p}.
Using both terms of Eq. (\ref{eq:e2.8}), instead, we obtain the spectrum
shown in Fig. \ref{fig:e2.2}, comparing much better with experiments.
The large influence of the non-local pseudopotential commutator on $d\rightarrow\;s/p$
optical transitions
can be understood by writing explicity the contribution of the second term of 
Eq. (\ref{eq:e2.8}):
\begin{multline}
\la \phi_{\(s/p\){\bf k}}|\[{\bf r},V_{NL}\]|\phi_{\(d\) {\bf k}}\ra= 
 \sum_{l=s,p,d} \iint d\,{\bf r}d\,{\bf r'} \phi^*_{\(s/p\){\bf k}}\({\bf r}\)\\
 \[{\bf r}V^l_{NL}\({\bf r},{\bf r'}\)-V^l_{NL}\({\bf r},{\bf r'}\){\bf r'}\]
 \phi_{\(d\){\bf k}}\({\bf r'}\),
 \label{eq:e2.11}
\end{multline}
where $V^l_{NL}\({\bf r},{\bf r'}\)$ is the l-orbital component of the pseudopotential,
and $\phi_{\(s/p\){\bf k}}\({\bf r}\)$, $\phi_{\(d\){\bf k}}\({\bf r}\)$ are the
$s/p$ like and $d$ like Bloch functions.
Now we can approximate the sum in Eq. (\ref{eq:e2.11}) with the leading terms,
to obtain:
\begin{multline}
\la \phi_{\(s/p\){\bf k}}|\[{\bf r},V_{NL}\]|\phi_{\(d\) {\bf k}}\ra\approx 
 \iint d\,{\bf r}d\,{\bf r'} \\
 \[\phi^*_{\(s/p\){\bf k}}\({\bf r}\)
 {\bf r}V^d_{NL}\({\bf r},{\bf r'}\)\phi_{\(d\){\bf k}}\({\bf r'}\)-\right.\\\left.
 \phi^*_{\(s/p\){\bf k}}\({\bf r}\)\(V^s_{NL}\({\bf r},{\bf r'}\)+V^p_{NL}\({\bf r},{\bf r'}\)\)
 {\bf r'}\phi_{\(d\){\bf k}}\({\bf r'}\)\].
 \label{eq:e2.12}
\end{multline}
In the case of copper the $d$ components of the pseudopotential differ from the $p$ component near
the origin by about 20 Hartrees, and this explain the strong influence on 
$\gee^{''}_M\(\go\)$.

Despite the strong improvement of the agreement with experiment obtained in Fig. \ref{fig:e2.2},
the theoretical curve exhibits an amplitude overstimation of about  20\%
with respect to the experimental absorption spectrum. 
\begin{figure}[H]
\begin{center}
\epsfig{figure=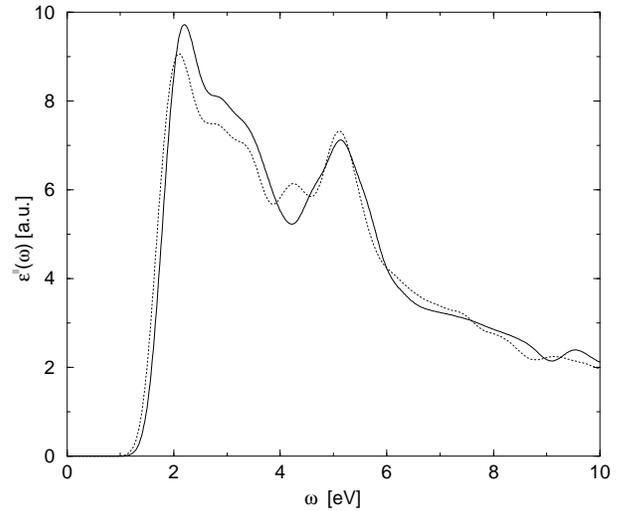,width=8cm}
\end{center}
\caption {\footnotesize{Effect of the use of a pseudopotential  including the 3s and 3p shells
in the valence on the calculated  $\varepsilon^{''} (\omega)$ for bulk Cu
(full line : ``3d'' PP; dotted line: ``3s'' PP. See text).
Due to the large number of plane-waves required by the ``3s'' PP, the comparison
is done using a small number of k-points and a relatively
large Gaussian broadening (N$_k=3,000$, $\gamma$ = 0.2 eV).
Local-Field effects are neglected, and Eq. (\ref{eq:e2.8}) is used.}}
\label{fig:e2.3}
\end{figure}

This drawback must be analyzed taking into account both the {\it physical approximations} involved
in our theoretical approach, and the possible {\it residual errors} due to the PP scheme.
Concerning the first ones, the most important point is the neglection of self--energy
effects in the bandstructure calculation, and of excitonic effects in the absorption process.
Concerning the PP scheme, instead, a possible reason for the overstimation of the spectrum intensity
could be related to the use of pseudo--wavefunctions instead of the all--electrons ones in Eq. 
(\ref{eq:e2.4}). This effect has been studied in atoms by the authors of Ref. \cite{read}: they 
found that PP calculations,
even when the second term of rhs of Eq. (\ref{eq:e2.8}) is correctly taken into account, 
can be affected by a small residual error due to the difference between all-electron wavefunctions and
pseudo wavefunctions {\it inside the core region}.
In the case of the Cu atoms, $3d\rightarrow 4p$ transitions were found to yield a matrix
element which was too large by about 10\%\cite{read}.

The effect of this overstimation of the $3d\rightarrow 4p$ intra-atomic optical matrix elements
on the calculated bulk spectrum is, however, not obvious. To clarify this point, we have performed
an accurate comparison of the results obtained by using the ``3d'' and  ``3s'' pseudopotentials,
both in case of the bulk crystal and for the isolated Cu atom.
For the latter case, our results are summarized in Table \ref{tab:e2.1}: while our ``3d''
pseudopotential gives about the same results as those obtained in Ref. \cite{read}, the ``3s''
PP reduces the error on the intra-atomic optical matrix
 elements to less than 1.5 \%.
However in the case of the 
bulk crystal the amplitude of our calculated spectrum does not change appreciably
when results obtained with ``3d'' and  ``3s'' pseudopotentials are compared 
(see Fig. \ref{fig:e2.3}).
This suggests that the overall intensity overestimation cannot be ascribed to the use of PP
wavefunctions in Eq. (\ref{eq:e2.4}).

The reason for this discrepancy between theory and experiment should hence be searched for within the 
physical approximations made, such as the neglection of self-energy effects that
are currently under investigation and will be the subject of a forthcoming 
paper\cite{gwcu}.

%%%%%%%%%%%%%%%%%%%%%%%%%%%%%%%%%%
\subsection{Intraband transitions}
\label{sec:drude}
%%%%%%%%%%%%%%%%%%%%%%%%%%%%%%%%%%

A peculiar characteristic of metals is the intra-band contribution to  the dielectric function, 
which, neglecting local field effects, is given by:
\begin{multline}
 \gee^{'}_{intra}\(\go\)\equiv 1-\lim_{{\bf q}\rightarrow {\bf 0}} 
 \left\{ \frac{4 \pi}{|{\bf q}|^2} 
 \sum_n
 \int_{BZ} \frac{d^3 {\bf k}}{\(2 \pi\)^3}\capo
 \[ f_n\({\bf k-q}\)-f_n\({\bf k}\)\] 
 \frac{|\la n{\bf k}|e^{i{\bf q}\cdot{\bf r}}|n {\bf k-q}\ra|^2}
 {\go+\gee_n\({\bf k-q}\)-\gee_n\({\bf k}\)}\right\},\label{eq:drude1}
\end{multline}
with the n-sum restricted only to semi-occupied bands. In the case of copper only the sixth
band contributes to Eq. (\ref{eq:drude1}).Indicating this band with $n_F$ and
using the relation
\begin{multline}
 f_{n_F}\({\bf k-q}\)-f_{n_F}\({\bf k}\)=\(f_{n_F}\({\bf k-q}\)-f_{n_F}\({\bf k}\)\)\\
 \{\theta\[ f_{n_F}\({\bf k-q}\)-f_{n_F}\({\bf k}\)\]-
 \theta\[ f_{n_F}\({\bf k}\)-f_{n_F}\({\bf k-q}\)\]\},
\label{eq:drude2}
\end{multline}
time-reversal symmetry allows one to rewrite Eq. (\ref{eq:drude1}) as
\begin{multline}
 \gee^{'}_{intra}\(\go\)\equiv 1-\lim_{{\bf q}\rightarrow {\bf 0}} 
 \left\{ \frac{8\pi}{|{\bf q}|^2}
 \int_{BZ} \frac{d^3 {\bf k}}{\(2 \pi\)^3}\capo
 \(f_{n_F}\({\bf k-q}\)-f_{n_F}\({\bf k}\)\)\theta\[ f_{n_F}\({\bf k-q}\)-f_{n_F}\({\bf k}\)\]\capo
 \frac{|\la n{\bf k}|e^{i{\bf q}\cdot{\bf r}}|n {\bf k-q}\ra|^2
 \(\gee_{n_F}\({\bf k}\)-\gee_{n_F}\({\bf k-q}\)\)}
 {\go^2-\(\gee_{n_F}\({\bf k}\)-\gee_{n_F}\({\bf k-q}\)\)^2}\right\},
\label{eq:drude3}
\end{multline}
In the  small ${\bf q}$ limit, Eq. (\ref{eq:drude3}) yields the
well-known Drude contribution to the 
dielectric function
\begin{align}
 \gee^{'}_{intra}\(\go\)\equiv 1-
 \frac{\go_D^2}{\go^2}+O\(q^2\)\label{eq:drude4},
\end{align}
with
\begin{multline}
 \go_D^2=\lim_{{\bf q}\rightarrow {\bf 0}} 
 \left\{ \frac{8 \pi}{|{\bf q}|^2}
 \int_{BZ} \frac{d^3 {\bf k}}{\(2 \pi\)^3}\capo
 \(f_{n_F}\({\bf k-q}\)-f_{n_F}\({\bf k}\)\)\theta\[ f_{n_F}\({\bf k-q}\)-f_{n_F}\({\bf k}\)\]\capo
 |\la n{\bf k}|e^{i{\bf q}\cdot{\bf r}}|n {\bf k-q}\ra|^2
 \(\gee_{n_F}\({\bf k}\)-\gee_{n_F}\({\bf k-q}\)\) \right\}.\label{eq:drude5}
\end{multline}
Since the theta function in Eq. (\ref{eq:drude5}) limits strongly the
region of BZ that contributes to the integral, the ${\bf k}$-space sampling   
and the modulus of the chosen ${\bf q}$ vector used in the numerical evaluation of
Eq. (\ref{eq:drude5})
become two critical convergence parameters: 
for a small $|{\bf q}|$ very few k-points will satisfy the condition 
$\[f_{n_F}\({\bf k-q}\)-f_{n_F}\({\bf k}\)\]\neq 0$.
In practice, $|{\bf q}|$ must be small enough to reproduce the ${\bf q}\rightarrow {\bf 0}$ limit in 
Eq. (\ref{eq:drude5}), but large enough to allow a suitable number of {\bf k}-points
to contribute to the sum.
To obtain a well converged $\go_D^2$ we found it necessary to use $\approx 16,000$ random {\bf k}-points in 
a region of the BZ such that
$\gee_{n_F}\({\bf k}\)$ is contained within $\gee_{Fermi}\pm 0.1$ eV.
The value  used was $|{\bf q}|=0.005\;a.u.$.
A fictious electronic temperature was introduced to smear out the Fermi surface, increasing
the number of $\({\bf k},{\bf k-q}\)$ pairs giving non zero contributions to Eq. (\ref{eq:drude5}).
In Table \ref{tab:drude1} we present our results as a function of the fictious electronic temperature.
The optical mass, defined as
\begin{align}
 m_{opt}=\(\frac{\go_c}{\go_D}\)^2\qquad\text{with}\quad 
 \go_c=\sqrt{\frac{4 \pi}{\Omega}}= 10.8\,\text{eV}
\end{align}
with $\Omega$ direct lattice cell volume,
converges to 1.36, a value in good agreement with the experiment\cite{phillipp}.

%%%%%%%%%%%%%%%%%%%%%%%%%%%%%%%%
\subsection{Local field effects}
%%%%%%%%%%%%%%%%%%%%%%%%%%%%%%%%

Local-Field effects are accounted for when the macroscopic dielectric function is
computed according to Eq. (\ref{eq:e2.1}--\ref{eq:e2.3}), i.e. by explicitly obtaining
$\chi_{{\bf G\;G'}}\({\bf q},\go\)$ as
\begin{multline}
\chi_{{\bf G\;G'}}\({\bf q},\go\)=
\chi_{{\bf G\;G'}}^0\({\bf q},\go\)\\ 
\[1-\frac{4 \pi}{|{\bf q+G}|^2} 
\chi^0_{{\bf G\;G'}}\({\bf q},\go\)\]^{-1},
\label{eq:lf1}
\end{multline}
We divide explicitly $\chi^0_{{\bf G\;G'}}\({\bf q},\go\)$ 
into intraband and interband contributions:
\begin{align}
 \chi^0_{{\bf G\;G'}}\({\bf q},\go\)=\chi^{inter}_{{\bf G\;G'}}\({\bf q},\go\)+
 \chi^{intra}_{{\bf G\;G'}}\({\bf q},\go\),
\label{eq:lf2}
\end{align}
where the interband part is given by Eq. (\ref{eq:e2.4}), while for the intraband 
contribution we have to evaluate:
\begin{multline}
 \chi^{intra}_{{\bf G\;G'}}\({\bf q},\go\)=\frac{1}{2} 
 \int_{BZ} \frac{d^3 {\bf k}}{\(2 \pi\)^3}\\
 \la n_F{\bf k-q}|e^{-i\({\bf q+G}\)\cdot{\bf r}}|n_F {\bf k}\ra
 \la n_F{\bf k}|e^{i\({\bf q+G'}\)\cdot{\bf r}'}|n_F {\bf k-q}\ra\\
 {\cal G}^0_{he}\(n_F,n_F,{\bf k},{\bf q},\go\).
\label{eq:lf3}
\end{multline}
with  ${\cal G}^0_{he}\(n,n',{\bf k},{\bf q},\go\)$ defined in Eq. (\ref{eq:e2.5}).
For $G\;,G'\neq\,0$ the oscillator
strength is calculated using Fast Fourier Transforms (FFT)
\begin{align}
 \la n{\bf k-q}|e^{-i\({\bf q+G}\)\cdot{\bf r}}|n' {\bf k}\ra=
 \la u_{n{\bf k}}|e^{-i{\bf G}\cdot{\bf r}}| u_{n'{\bf k}}\ra+O\(q\),
\end{align}
while  for the $G = G' = 0$ element of $\chi^{inter}$ we use Eq. \ref{eq:e2.7}.
\begin{figure}[H]
\begin{center}
\epsfig{figure=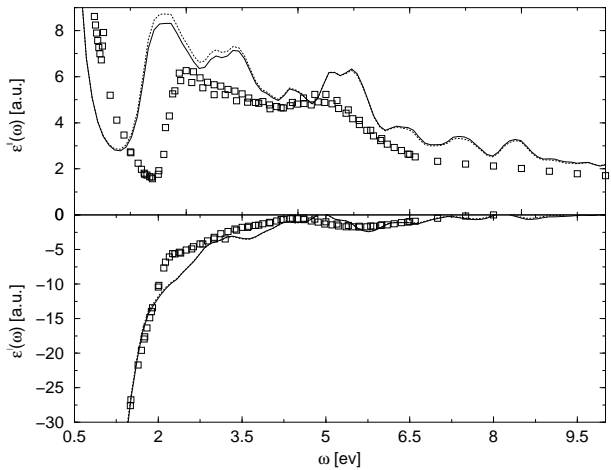,width=8cm}
\end{center}
\caption {\footnotesize{Real and imaginary part of $\varepsilon_M (\omega)$ for bulk Cu
with (full line) and without (dotted line) inclusion of Local-Field effects,
compared with experimental data (squares) from Ref.{\protect\cite{palik}}.
Theoretical spectra are computed with N$_k=110$ (in the irreducible wedge of BZ),
and a Lorentzian broadening of 0.4 eV (see text).}}
\label{fig:lf.1}
\end{figure}

The method presented in Section \ref{sec:drude} for the calculation of $\go_D$ could,
in principle, be extended to $\chi^{intra}_{{\bf G\;G'}}\({\bf q},\go\)$.
Unfortunately the explicit calculation of $\chi^{intra}$ for all the
$\({\bf G,G'}\)$ pairs is computationally prohibitive, due to the large number of
{\bf k}-points required to reach convergence.
To overcome these difficulties, we have evaluated Eq. (\ref{eq:lf3}) on a limited 
number of {\bf k}-points (a
MP grid of 110 points  in the irreducible wedge).
${\cal G}^0_{he}\(n_F,n_F,{\bf k},{\bf q},\go\)$ is different from zero only
for ${\bf k}$ very close to the Fermi Surface (which, in the grid, 
coincides with one particular point 
${\bf k}_F$) where $f_{n_{F}}\({\bf k-q}\)\(2-f_{n_{F}}\({\bf k}\)\)\neq 0$. 
The oscillator strengths can be considered to be almost constant in the vicinity of the Fermi Surface.
The same ansatz cannot be applied to ${\cal G}^0_{he}\(n_F,n_F,{\bf k},{\bf q},\go\)$;
however we can use 
the property that near the Fermi surface the metallic band dispersion of copper is well approximated by
a free-metal one, and make the following assumption for $\chi^{intra}_{{\bf G\;G'}}\({\bf q},\go\)$:
\begin{multline}
\chi^{intra}_{{\bf G\;G'}}\({\bf q},\go\)\approx\\
 \frac{1}{N_s}
 \[\sum_{{\bf R}}
 \la n_F{\bf Rk_F-q}|e^{-i\({\bf q+G}\)\cdot{\bf r}}|n_F {\bf Rk_F}\ra \capo
 \la n_F{\bf Rk_F}|e^{i\({\bf q+G'}\)\cdot{\bf r}'}|n_F {\bf Rk_F-q}\ra\]
 \pi_0\({\bf q},\go\),
\end{multline}
where ${\bf R}$ is one of the $N_s$ symmetry operations not in the point group of ${\bf k}_F$.
$\pi_0\({\bf q},\go\)$ is the non-interacting polarization calculated for a
jellium model\cite{mattuck} with a density $n_{el}$ yielding a classic plasma frequency 
$\go_p=\sqrt{4\pi n_{el}}=9.27$ eV,
corresponding to the value of $\go_D$ calculated in Section \ref{sec:drude}.
\begin{multline}
 \pi_0\({\bf q},\go\)=-\frac{1}{2 \pi^2 |{\bf q}|}\int_0^{k^{jel}_F |{\bf q}|} x\,dx\\
 \(\frac{1}{\go-k^{jel}_F |{\bf q}|+ i\eta}-\frac{1}{\go-k^{jel}_F |{\bf q}|- i\eta}\).
\end{multline}
where
\begin{align}
x=\frac{\go}{k^{jel}_F |{\bf q}|}.
\end{align}
and
\begin{align}
k^{jel}_F=\(r\pi^2 n_{el}\)^{1/3}.
\end{align}
\begin{figure}[H]
\begin{center}
\epsfig{figure=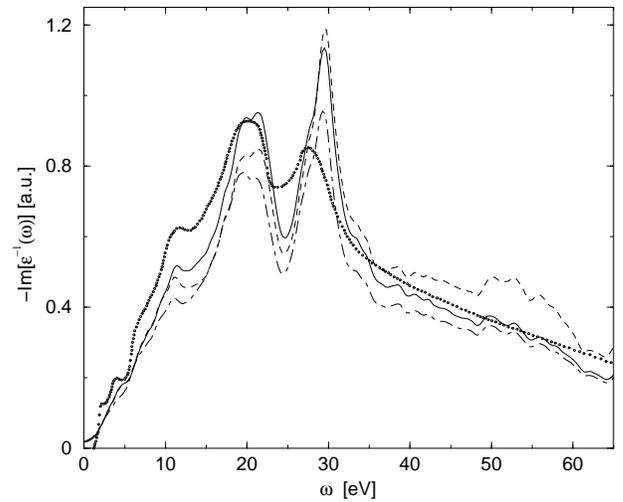,width=8cm}
\end{center}
\caption{\footnotesize{Imaginary part of the inverse
macroscopic dielectric function of Cu including  Local-Field effects
and the Drude contribution (full line).
Dashed line: results without Local-Field effects;
Long-short dashes: with Local-Field effects, but without the  Drude contribution.
All theoretical spectra are computed with the same N$_k$ of Fig. \ref{fig:lf.1} but
with a  Lorentzian broadening of  0.9 eV.
Points are the EELS data from Ref.{\protect \cite{palik}}.}}
\label{fig:lf.2}
\end{figure}
In Fig. \ref{fig:lf.1} we compare our results for $\varepsilon_M^{''}(\omega)$
with and without Local Field Effects. The differences are small, consistently with the 
fact that, as expected, no large LFE are present in a metal. 
However Local Field effects are more important on the EELS spectrum,
 as shown in Fig. \ref{fig:lf.2}:
in the high energy region the full inversion of $\chi_{{\bf G\;G'}}\({\bf q},\go\)$ matrix
corrects an overstimation of the intensity.
The inclusion of intraband transitions, on the other hand, turns out to be necessary
not only to describe correctly the behaviour of $\varepsilon^{''}$ at low frequencies, 
but also to improve the agreement of the calculated $\varepsilon ^{-1}(\omega)$ with
the experimental EELS data (Fig.\ref{fig:lf.2}).

%%%%%%%%%%%%%%%%%%%%%
\section{Conclusions}
%%%%%%%%%%%%%%%%%%%%%

Several conclusions can be drawn from the presented results. First, we have shown 
that fully converged DFT-LDA calculations for bulk copper using plane waves can be 
performed without requiring an exceedingly large basis set, by using soft 
norm-conserving pseudopotentials including the 3d electrons in the valence. 

The strong non-locality implied by such pseudopotentials reflects itself in a very 
large contribution of the commutator between the PP and  the position operator in the  
calculation of the matrix elements entering the dielectric function. 

Similar calculations including also the 3s and 3p shells in the valence are also 
feasible using plane waves, but show very little difference on the bandstructure 
and on the optical matrix elements not directly involving those shells.

At difference with the case of simple semiconductors, the disagreement between the 
LDA theoretical bandstructure and the experimental one cannot be corrected 
by a rigid shift of the Kohn-Sham eigenvalues: this suggests the importance of 
many body effects in the quasiparticle spectrum of Cu. These effects should be taken into 
account through a calculation including self-energy contributions 
\cite{gwcu}. 
By contrast, the computed optical mass, which involves  only 
intraband transitions at the Fermi energy, is in very good agreement with the 
experimental data. 
Also the theoretical absorption spectrum, where only LDA eigenvalues within a 
few eV around the Fermi energy are important, displays  a quite satisfactory 
agreement with the experimental one. The overall overestimation of about 20 
percent in the amplitude of the main peaks cannot be attributed to 
the effect of wavefunction pseudization {\it inside} the core region,
and should be ascribed to the neglection of self-energy (and excitonic) effects.
Finally, local field effects on the macroscopic dielectric function turn out to be 
negligible, with or without the inclusion of the Drude term.

%%%%%%%%%%%%%%%%%%%%%%%%%%
\section*{Acknowledgments}
%%%%%%%%%%%%%%%%%%%%%%%%%%
This work has been supported by 
the INFM  PRA project "1MESS" and 
by the EU through the NANOPHASE Research Training
Network (Contract No. HPRN-CT-2000-00167.
We thank L. Reining and A. Rubio for useful discussions.
We are grateful to  S. Goedecker for providing us an efficient
code for Fast Fourier Transforms\cite{goedecker}.

\end{multicols}
\widetext

%%%%%%%%%%%%%%%%%%%%%%%%%% TABLES %%%%%%%%%%%%%%%%%%%%%%%%%%%%%%%%

\begin{table}
\begin{tabular}{ccccc} 
Pseudopotential & r$_s$ [Bohr] & r$_p$ [Bohr] & r$_d$ [Bohr] & Reference l--component\\ 
\hline
``3d''         &  1.19 (H)                  &  1.19 (H)       &  2.08 (TM) & s   \\
``3d+NLCC''    &  1.10 (H)                  &  1.19 (H)       &  2.08 (TM) & s   \\
``3s''         &  0.49 (H)                  &  0.60 (H)       &  1.19 (TM) & p   \\
\end{tabular}
\caption{Cu Pseudopotential cutoff radii and reference components. (H) and (TM)
stand for Hamann (ref.  {\protect\cite{hamann}}) and 
Troulliers--Martins (ref.  {\protect\cite{TM}}) schemes, respectively.
All our pseudopotentials  are norm--conserving in the sense
of Bachelet, Hamann and Schl\"{u}ter{\protect \cite{BHS}}, and have been produced 
using  a publicly available Fortran  code  {\protect\cite{fhi96md}}. They are also
available in numerical form upon request by e-mail to: onida@roma2.infn.it}
\label{table_rc_pseu}
\end{table}

\begin{table}
\begin{tabular}{cccc} 
PW cutoff (Ry) & Nk & a$_0$ [Bohr] & B$_0$ [MBar] \\
\hline
75        &  10              & 6.77    & 1.73  \\
75        &  28              & 6.73    & 1.79  \\
75        &  60              & 6.74    & 1.83  \\
60        &  28              & 6.73    & 1.83  \\
100+NLCC  &  28              & 6.79    & 1.71  \\
Expt.     &  --              & 6.82    & 1.83  \\
\end{tabular}
\caption{ Convergence of the calculated ground--state properties of bulk Cu.}
\label{table_conv}
\end{table}

\begin{table}
\begin{tabular}{lccc} 
 &   & Experiment {\protect\cite{courths}} & Present work \\ \hline
   & $\Gamma_{12}$ & $-2.78$ & $-2.33$ \\
positions of d-bands   & $X_5$ & $-2.01$ & $-1.46$ \\
   & $L_3$ & $-2.25$ & $-1.69$ \\ \hline
 	& $\Gamma_{12}-\Gamma_{25'}$ & $0.81$ & $0.91$ \\
	& $X_5-X_3$ & $2.79$ & $3.23$ \\
widths of d-bands & $X_5-X_1$ & $3.17$ & $3.70$ \\
	& $L_3-L_3$ & $1.37$ & $1.58$ \\ 
	& $L_3-L_1$ & $2.91$ & $3.72$ \\ \hline
positions of sp-bands & $\Gamma_{1}$ & $-8.60$ & $-9.85$ \\
 & $L_{2'}$ & $-0.85$ & $-1.12$ \\ \hline
L-gap & $L_1-L_{2'}$ & $4.95$ & $4.21$ \\
\end{tabular}
\caption{Comparison of Cu band widths and energy position
with experimental values {\protect\cite{courths}} at high-symmetry points. All energies
in eV.}
\label{tab:lda_bands.1}
\end{table}

\begin{table}
\begin{tabular}{cccc}
Optical transitions & {\it AE} & ''3d`` PP& ''3s`` PP \\
\hline
$4s\;\rightarrow\;4p$ & $1.732$ & $1.738$ & $1.713$ \\
$3d\;\rightarrow\;4p$ & $0.406$ & $0.453$ & $0.412$ \\
\end{tabular}
\caption{Optical matrix elements for the Cu atom,
calculated in the All-electrons (AE) scheme, and with
the pseudopotentials used in the present work.
Values are in atomic units.}
\label{tab:e2.1}
\end{table}

\begin{table}
\begin{tabular}{ccccc} 
$T_{el}\;[eV]$ & Number of intraband transitions & $\go_D$ [ev] & $m_{opt}$ \\ 
\hline
0.1            &  14,617                         &  9.39         &  1.32    \\
0.01           &  2,456                          &  9.41         &  1.32    \\
0.001          &  534                            &  9.25         &  1.36    \\
0.0001         &  310                            &  9.28         &  1.36    \\
0.00001        &  296                            &  9.27         &  1.36    \\
\end{tabular}
\caption{Cu Drude plasma frequency and optical mass values obtained using
different fictious electronic temperatures; the experimental value 
for  $m_{opt}$ is $1.35${\protect\cite{phillipp}}.}
\label{tab:drude1}
\end{table}


\begin{references}
\bibitem{courths}	R. Courths and S. H\"{u}fner,
			Phys. Rep. {\bf 112}, 53 (1984).
\bibitem{campillo} 	I. Campillo, A. Rubio and J.M. Pitarke, 
            		Phys. Rev. B {\bf 59}, 12188 (1999).

\bibitem{vdb}           D. Vanderbilt, Phys. Rev. B {\bf 41}, 
                        7892 (1990).

\bibitem{rappe}         A. M. Rappe, K. M. Rabe, E. Kaxiras, 
                        and J.D. Johannopoulos, Phys. Rev.
                        B {\bf 41}, 1227 (1990).

\bibitem{shir}          E.L. Shirley, D.C. Allan, R.M. Martin,
                        and J. D. Joannopoulos, Phys. Rev. 
                        B {\bf 40}, 3652 (1989).

\bibitem{TM} 		N. Troullier and J.L. Martins, 
            		Phys. Rev. B {\bf 43}, 1993 (1991).
\bibitem{BHS}		G. Bachelet, D.R. Hamann and M. Schl\"{u}ter
	 		Phys. Rev. B {\bf 26}, 4199 (1982).
\bibitem{hamann}	D. R. Hamann, 
			Phys. Rev. B {\bf 40}, 2980 (1989).
\bibitem{gonze_ghost}	X. Gonze, P. K\"{a}ckell, M. Scheffler, 
			Phys. Rev. B {\bf 41}, 12264 (1990).
\bibitem{KB}		l. Kleinman and D. M. Bylander,
	 		Phys. Rev. Lett. {\bf 48}, 1425 (1982).
\bibitem{sno}		M. Meyer, G. Onida, M. Palummo and L. Reining,
                        Phys. Rev. B 64, 045119 (2001).
\bibitem{NLCC}		S. G. Louie, S. Froyen and M. L. Cohen,
            		Phys. Rev. B {\bf 26}, 1738 (1982).
\bibitem{CP}		R. Car and M. Parrinello,
	 		Phys. Rev. Lett. {\bf 55}, 2471 (1985).
\bibitem{cepal}		D.M. Ceperley and B.I. Alder, 
			Phys. Rev. Lett. {\bf 45}, 566 (1980).
\bibitem{perzun}	J. P. Perdew and A. Zunger,
            		Phys. Rev. B {\bf 23}, 5048 (1981).
\bibitem{hedin_lund}	L. Hedin and B. I. Lundqvist,
            		J. Phys. C. {\bf 4}, 2064 (1971).
\bibitem{MP}		H. J. Monkorst and J. D. Pack,
            		Phys. Rev. B {\bf 13}, 5188 (1976).
\bibitem {nota1}
 However, as long as we are interested in bare LDA result,
 the  overall effect  on the 3d - 4s bandstructure  and on the absorption 
 spectrum is  small and, a posteriori, we can simply use our ``3d'' PP.
\bibitem{gunn_rev}	R. O. Jones and O. Gunnarsson,
			Rev. Mod. Phys. {\bf 61}, 689 (1989).
\bibitem{REPLY_ALBRECHT}	S. Albrecht, L. Reining, G. Onida, 
                                V. Olevano and R. Del Sole, 
	 			Phys. Rev. Lett. {\bf 83}, 3971 (1999).
\bibitem{rodolfo1} 	R. Del Sole and R. Girlanda, 
            		Phys. Rev. B {\bf 48}, 11789 (1993).
\bibitem{hybertsen1} 	M. S. Hybertsen and S. G. Louie, 	
            		Phys. Rev. B {\bf 35}, 5585 (1987).
\bibitem{read}  	J. Read and R.J. Needs, 
			Phys. Rev. B {\bf 44}, 13071 (1991).
\bibitem{gwcu}  	A. Marini, R. Del Sole and G. Onida, unpublished. 
\bibitem{phillipp}	H. Ehrenreich and H.R. Phillipp,
			Phys. Rev. {\bf 128}, 1622 (1962).
\bibitem{mattuck}       R. D. Mattuck, {\em A guide to Feynman diagrams in the
                        Many-Body problem}, McGraw-Hill, New York (1976)
                        p. 196-200.
\bibitem{goedecker}	S. Goedecker,
			Comp. Phys. Commun. {\bf 76}, 294 (1993);
			SIAM Journal on Scientific Computing {\bf 18}, 1605 (1997).
\bibitem{fhi96md}	M. Bockstedte, A. Kley, J. Neugebauer, M. Scheffler,
			Comput. Phys. Comm. {\bf 107}, 187 (1997).
\bibitem{expt}		R.W.G. Wyckoff, {\it Crystal Structure}, 2nd ed., Interscience,
                	New York (1976).
\bibitem{palik}		E.D. Palik, {\it Handbook of Optical Constatnts
	 		of solids} (Academic Press, New York, 1985).
\end{references}
\end{document}